\documentclass[11pt]{article}

\usepackage{a4wide}
\usepackage{multirow}
\usepackage{endfloat}
\usepackage{graphicx}
\usepackage{amsmath}
\usepackage{amssymb}
\usepackage{amsfonts}
\usepackage{natbib}
\usepackage{pstricks}
\usepackage{psfrag}
\usepackage{setspace} 
\usepackage{aeguill} 

\def\ee{\mathbb{E}} %

\def\bX{\mbox{\boldmath $X$}}
\def\bu{\mbox{\boldmath $u$}}

\begin{document} 

\title{\bf Global sensitivity analysis of computer models with functional inputs} 

\author{\bf Bertrand IOOSS$^\ast$ and Mathieu RIBATET$^\dag$}
\date{}

\maketitle 

\begin{center}
  Submitted to: {\it Reliability Engineering and System Safety}\\
  for the special SAMO 2007 issue

  \vspace{0.2cm}

  $^\ast$ CEA, DEN, DER/SESI/LCFR, F-13108 Saint-Paul-lez-Durance, 
  France.

  $^\dag$ \'Ecole Polytechnique F\'ed\'erale de Lausanne, Chair of
  Statistics, STAT-IMA-FSB-EPFL, Station 8, CH-1015 Lausanne,
  Switzerland.

  \vspace{0.2cm}
  Corresponding author: B. Iooss ; Email: bertrand.iooss@cea.fr\\
  Phone: +33 (0)4 42 25 72 73 ; Fax: +33 (0)4 42 25 24 08
\end{center}

\doublespacing

\abstract{ Global sensitivity analysis is used to quantify the
  influence of uncertain input parameters on the response variability
  of a numerical model.  The common quantitative methods are
  appropriate with computer codes having scalar input variables.  This
  paper aims at illustrating different variance-based sensitivity
  analysis techniques, based on the so-called Sobol's indices, when
  some input variables are functional, such as stochastic processes or
  random spatial fields.  In this work, we focus on large cpu time
  computer codes which need a preliminary metamodeling step before
  performing the sensitivity analysis.  We propose the use of the
  joint modeling approach, i.e., modeling simultaneously the mean and
  the dispersion of the code outputs using two interlinked Generalized
  Linear Models (GLM) or Generalized Additive Models (GAM).  The
  ``mean model'' allows to estimate the sensitivity indices of each
  scalar input variables, while the ``dispersion model'' allows to
  derive the total sensitivity index of the functional input
  variables.  The proposed approach is compared to some classical 
  sensitivity analysis
  methodologies on an analytical function.  Lastly, the new
  methodology is applied to an industrial computer code that simulates
  the nuclear fuel irradiation.}


\section{INTRODUCTION}

Modern computer codes that simulate physical phenomenas often take as
inputs a high number of numerical parameters and physical variables,
and return several outputs - scalars or functions. For the development
and the use of such computer models, Sensitivity Analysis (SA) is an
invaluable tool.  The original technique, based on the derivative
computations of the model outputs with respect to the model inputs,
suffers from strong limitations for computer models simulating
non-linear phenomena.  More recent global SA techniques take into
account the entire range of variation of the inputs and aim to
apportion the whole output uncertainty to the input factor
uncertainties (Saltelli et al.  \cite{salcha00}). The global SA
methods can also be used for model calibration, model validation,
decision making process, i.e., any process where it is useful to know
which are the variables that mostly contribute to the output
variability.

The common quantitative methods are applicable to computer codes with
scalar input variables.  For example, in the nuclear engineering
domain, global SA tools have been applied to numerous models where all
the uncertain input parameters are modeled by random variables,
possibly correlated - such as thermal-hydraulic system codes (Marquès
et al. \cite{marpig05}), waste storage safety studies (Helton et al.
\cite{heljoh06}), environmental model of dose calculations (Iooss et
al. \cite{ioovan06}), reactor dosimetry processes (Jacques et al.
\cite{jaclav06}).  Recent research papers have tried to consider more
complex input variables in the global SA process, especially in
petroleum and environmental studies:
\begin{itemize}
\item Tarantola et al. \cite{targig02} work on an environmental
  assessment on soil models that use spatially distributed maps
  affected by random errors.  This kind of uncertainty is modeled by a
  spatial random field (following a specified probability
  distribution), simulated at each code run.  For the SA, the authors
  propose to replace the spatial input parameter by a ``trigger''
  random parameter $\xi$ that governs the random field simulation. For
  some values of $\xi$, the random field is simulated and for the
  other values, the random field values are put to zero.  Therefore,
  the sensitivity index of $\xi$ is used to quantify the influence of
  the spatial input parameter.
\item Ruffo et al. \cite{rufbaz06} evaluate an oil reservoir
  production using a model that depends on different heterogeneous
  geological media scenarios. These scenarios, which are of limited
  number, are then substituted for a discrete factor (a scenario
  number) before performing the SA.
\item Iooss et al. \cite{ioorib07} study a groundwater radionuclide
  migration model which depend on several random scalar parameters and
  on a spatial random field (a geostatistical simulation of the
  hydrogeological layer heterogeneity).  The authors propose to
  consider the spatial input parameter as an ``uncontrollable''
  parameter. Therefore, they fit on a few simulation results of the
  computer model a double model, called a joint model: the first
  component models the effects of the scalar parameters while the
  second models the effects of the ``uncontrollable'' parameter.
\end{itemize}

In this paper, we tackle the problem of the global SA for numerical
models and when some input parameters $\varepsilon$ are functional.
$\varepsilon(\bu)$ is a one or multi-dimensional stochastic function
where $\bu$ can be spatial coordinates, time scale or any other
physical parameters.  Our work focuses on models that depend on scalar
parameter vector $\bX$ and involve some stochastic process simulations
or random fields $\varepsilon(\bu)$ as input parameters.  The computer
code output $Y$ depends on the realizations of these random functions.
These models are typically non linear with strong interactions between
input parameters.  Therefore, we concentrate our methodology on the
variance based sensitivity indices estimation; that is, the so-called
Sobol's indices (Sobol \cite{sob01}, Saltelli et al. \cite{salcha00}).

To deal with this situation, a first natural approach consists in
using either all the discretized values of the input functional
parameter $\varepsilon(\bu)$ or its decomposition into an appropriate
basis of orthogonal functions.  Then, for all the new scalar
parameters related to
$\varepsilon(\bu)$, 
sensitivity indices are computed. However, in the case of complex
functional parameters, this approach seems to be rapidly intractable
as these parameters cannot be represented by a small number of scalar
parameters (Tarantola et al. \cite{targig02}).  Moreover, when dealing
with non physical parameters (for example coefficients of orthogonal
functions used in the decomposition), sensitivity indices
interpretation may be laborious.  Indeed, most often, physicists would
prefer to obtain one global sensitivity index related to
$\varepsilon(\bu)$.  Finally, a major drawback for the decomposition
approach is related to the uncertainty modeling stage. More precisely,
this approach needs to specify the probability density functions for
the coefficients of the decomposition.

The following section presents three different strategies to compute
the Sobol's indices with functional inputs: (a) the macroparameter
method, (b) the ``trigger'' parameter method and (c) the proposed
joint modeling approach.  Section~\ref{sec:analyticalExample} compares
the relevance of these three strategies on an analytical example: the
WN-Ishigami function.  Lastly, the proposed approach is illustrated on
an industrial computer code simulating fuel irradiation in a nuclear
reactor.

\section{COMPUTATIONAL METHODS OF SOBOL'S INDICES}

First, let us recall some basic notions about Sobol's indices.  Let
define the model

\begin{equation}
  \label{eqmodeldeterm}
  \begin{array}{rcl}
    f : & \mathbb{R}^p & \rightarrow \mathbb{R} \\
    & \bX & \mapsto Y=f(\bX)
  \end{array}
\end{equation}
where $Y$ is the code output, $\bX=(X_1,\ldots,X_p)$ are $p$
independent inputs, and $f$ is the model function.  $f$ is considered
as a ``black box'', i.e. a function whose analytical formulation is
unknown.  The main idea of the variance-based SA methods is to
evaluate how the variance of an input or a group of input parameters
contributes to the output variance of $f$.  These contributions are
described using the following sensitivity indices:

\begin{equation}
  \label{eqindordre1}
  S_i=\frac{\mbox{Var}\left[\mathbb{E}\left(Y|X_i\right)\right]}{\mbox{Var}(Y)},
  \quad  S_{ij} = \frac{\mbox{Var}\left[\mathbb{E}\left(Y|X_i
        X_j\right)\right]}{\mbox{Var}(Y)} - S_i -S_j,  \quad  S_{ijk} = \ldots
\end{equation} 
These coefficients, namely the Sobol's indices, can be used for any
complex model functions $f$.  The second order index $S_{ij}$
expresses the model sensitivity to the interaction between the
variables $X_i$ and $X_j$ (without the first order effects of $X_i$
and $X_j$), and so on for higher orders effects.  The interpretation
of these indices is natural as all indices lie in $[0,1]$ and their
sum is equal to one. The larger an index value is, the greater is
the importance of the variable or the group of variables related to
this index.

For a model with $p$ inputs, the number of Sobol's indices is $2^p-1$;
leading to an intractable number of indices as $p$ increases. Thus, to
express the overall output sensitivity to an input $X_i$, Homma \&
Saltelli \cite{homsal96} introduce the total sensitivity index:

\begin{eqnarray}\label{indtotal}
  S_{T_i} = S_i + \sum_{j \neq i} S_{ij} + \sum_{j \neq i, k \neq i,
    j<k} S_{ijk} + \ldots = \sum_{l \,\in \,\# i}S_l 
\end{eqnarray}
where $\# i$ represents all the ``non-ordered'' subsets of indices
containing index $i$.  Thus, $\sum_{l \,\in \,\# i}S_l$ is the sum of
all the sensitivity indices having $i$ in their index.  The estimation
of these indices (Eqs. (\ref{eqindordre1}) and (\ref{indtotal})) can
be performed by simple Monte-Carlo simulations based on independent samples
(Sobol \cite{sob93}, Saltelli \cite{sal02}), or by refined sampling
designs introduced to reduce the number of required model evaluations
significantly, for instance FAST (Saltelli et al.  \cite{saltar99})
and quasi-random designs (Saltelli et al.  \cite{salrat08}).

Let us now consider a supplementary input parameter which is a
functional input variable $\varepsilon(\bu) \in \mathbb{R}$ where $\bu
\in \mathbb{R}^d$ is a $d$-dimensional location vector.
$\varepsilon(\bu)$ is defined by all its marginal and joint
probability distributions. In this work, it is supposed that random
function realizations can be simulated.  For example, these
realizations can be produced using geostatistical simulations
(Lantuéjoul \cite{lan02}) or stochastic processes simulations (Gentle
\cite{gen03}).  Our model writes now 

\begin{equation}
  Y=f(\bX,\varepsilon)
\end{equation}
and in
addition to the Sobol's indices related to $\bX$, our goal is to derive
methods to compute the sensitivity indices relative to $\varepsilon$,
i.e., $S_\epsilon$ (first order index), $S_{T_\varepsilon}$ (total
sensitivity index), $S_{i\varepsilon}$ (second order indices),
$S_{ij\varepsilon}$, \ldots

\subsection{The macroparameter method}\label{secmacro}

With the macroparameter method, the functional input parameter is not
seen as a functional by the computer code.  It is discretized in a
potentially large number of values (for example several thousands),
each of them being an input scalar parameter of the computer code.  As
all these values come from the functional input parameter (which
posseses a specific correlation structure), they can be considered as
an ensemble of correlated input parameters.  Taking into account
correlation between input variables in sensitivity analysis has been a
challenging problem, recently solved by a few authors (see Da Veiga et
al. \cite{davwah08} for a recent review).

One solution, proposed by Jacques et al. \cite{jaclav06}, to deal with
correlated input parameters, is to consider multi-dimensional
sensitivity indices (Sobol \cite{sob01}): each group of correlated
parameters is considered as a multi-dimensional parameter or
macroparameter.  One therefore performs a sensitivity analysis by
groups of correlated parameters.  To estimate Sobol indices (first
order, second order, \ldots, total), a large number of input
parameters (correlated and non correlated) have to be generated.  As
we know how to generate independent samples of a correlated variables
group, the simple Monte-Carlo sampling technique can be used (Sobol
\cite{sob93}, Saltelli \cite{sal02}).  However, more efficient
techniques than simple Monte-Carlo (in terms of the required size
sample), as FAST or quasi Monte-Carlo which use deterministic samples,
are prohibited with correlated input variables.

In our context, this approach, using the simple Monte-Carlo algorithm,
seems to be relevant as the input functional parameter
$\varepsilon(\bu)$ can be considered as a single multi-dimensional
parameter (i.e. a macroparameter).  For instance, the first order
Sobol's index related to $\varepsilon(\bu)$ is defined as previously
by

\begin{eqnarray}\label{eqsoboleps}
  S_\varepsilon =
  \frac{\mbox{Var}\left[\mathbb{E}\left(Y|\varepsilon\right)\right]}{\mbox{Var}(Y)}  
\end{eqnarray} 
A simple way to estimate $S_\varepsilon = D_\varepsilon/D$ is based on
the Sobol \cite{sob93} algorithm:

\begin{subequations}
  \begin{eqnarray}
    \label{eqsobolalgo1}
    \hat{f}_0 & = & \displaystyle \frac{1}{N} \sum_{k=1}^N
    f(\bX_k^{(1)},\varepsilon_k)\\ 
    \label{eqsobolalgo2}
    \hat{D} & = & \displaystyle \frac{1}{N-1} \sum_{k=1}^N
    f^2(\bX_k^{(1)},\varepsilon_k) - \hat{f}_0^2\\ 
    \label{eqsobolalgo3}
    \hat{D}_\varepsilon & = & \displaystyle \frac{1}{N-1} \sum_{k=1}^N
    f(\bX_k^{(1)},\varepsilon_k) f(\bX_k^{(2)},\varepsilon_k) -
    \hat{f}_0^2
  \end{eqnarray}
\end{subequations}
where $(\bX_k^{(1)})_{k=1\ldots N}$ and $(\bX_k^{(2)})_{k=1\ldots N}$
are two independent sets of $N$ simulations of the input vector $\bX$
and $(\varepsilon_k)_{k=1\ldots N}$ is a sample of $N$ realizations of
the random function $\varepsilon(\bu)$.  To compute the sensitivity
indices $S_i$, the same algorithm is used with two independent samples
of $(\varepsilon_k)_{k=1\ldots N}$.  In the same way, the total
sensitivity index $S_{T_\varepsilon}$ is derived from the algorithm of
Saltelli \cite{sal02}.

The major drawback of this method is that it may be cpu time
consuming, mainly because of the sampling method.  If $d$ is the
number of indices to be estimated, the cost of the Sobol's algorithm
is $n=N(d+1)$ while the cost of Saltelli's algorithm to estimate $d$
first order and $d$ total sensitivity indices is $n=N(d+2)$.  It is
well known that, for complex computer models, an accurate estimation
of Sobol's indices by the simple Monte-Carlo method (independent random
samples) requires $N>1000$, i.e. more than thousand model evaluations
for one input parameter (Saltelli et al. \cite{salrat08}).  In complex
industrial applications, this approach is intractable due to the cpu
time cost of one model evaluation and the possible large number of
input parameters. 

\subsection{The ``trigger'' parameter method}

Dealing with spatially distributed input variables, Tarantola et al.
\cite{targig02} propose an alternative that uses an additional scalar
input parameter $\xi$ - called the ``trigger'' parameter. $\xi \sim
U[0,1]$ governs the random function simulation.  More precisely, for
each simulation, if $\xi < 0.5$, the functional parameter
$\varepsilon(\bu)$ is fixed to a nominal value $\varepsilon_0(\bu)$
(for example the mean $\ee[\varepsilon(\bu)]$), while if $\xi > 0.5$,
the functional parameter $\varepsilon(\bu)$ is simulated.  Using this
methodology, it is possible to estimate how sensitive the model output
is to the presence of the random function.  Tarantola et
al. \cite{targig02} use the Extended FAST method to compute the first
order and total sensitivity indices of $6$ scalar input factors and
$2$ additional ``trigger'' parameters. For their study, the
sensitivity indices according to the ``trigger'' parameters are small
and the authors conclude that it is unnecessary to model these spatial
errors more accurately.


Contrary to the previous method, there is no restriction about the
sensitivity indices estimation procedure - i.e. Monte-Carlo, FAST,
quasi Monte-Carlo.  However, there are two major drawbacks for this approach:
\begin{itemize}
\item As the macroparameter method, it also requires
  the use of the computer model to perform the SA and it may be
  problematic for large cpu time computer models.  
  This problem can be compensated by the use of an efficient quasi Monte-Carlo algorithm
  for which the sampling design size can be decreased to $N=100$.
\item As underlined by Tarantola et
  al.  \cite{targig02}, $\xi$ reflects only the presence or the absence
  of the stochastic errors on $\varepsilon_0(\bu)$.  Therefore, the term
  $\mbox{Var}[\ee(Y|\xi)]$ does not quantify the contribution
  of the random function variability to the output variability
  $\mbox{Var}(Y)$. We will discuss about the significance of 
  $\mbox{Var}[\ee(Y|\xi)]$ later, during our analytical function application.
\end{itemize}

\subsection{The joint modeling approach}

To perform a variance-based SA for time consuming computer models,
some authors propose to approximate the computer code, starting from
an initial small-size sampling design, by a mathematical function
often called response surface or metamodel (Marseguerra et
al. \cite{marmas03}, Volkova et al.  \cite{volioo08}, Fang et
al. \cite{fanli06}).  This metamodel, requiring negligible cpu time,
is then used to estimate Sobol's indices by any method, for example the
simple Monte-Carlo algorithm.  For metamodels with sufficient
prediction capabilities, the bias between the exact Sobol's indices
(from the computer code) and the Sobol's indices estimated via the
metamodel is negligible.  Indeed, it has been shown that the
unexplained variance part of the computer code by the metamodel (which
can be measured) corresponds to this bias (Sobol \cite{sob03}).
Several choices of metamodel can be found in the literature:
polynomials, splines, Gaussian processes, neural networks, \ldots\;
The fitting process is often based on least squares regression
techniques.  Thus, for the functional input problem, one strategy may
be to fit a metamodel with a multi-dimensional scalar parameters
representing $\varepsilon(\bu)$ as an input parameter - i.e. its
discretization or its decomposition into an appropriate basis.  This
process would correspond to a metamodeling approach for the
macroparameter method.  However, this approach seems to be
impracticable due to the potential large number of scalar parameters:
applying regression techniques supposes to have more observation
points (simulation sets) than input parameters and important numerical
problem (like matrix conditioning) might occur while dealing with
correlated input parameters.

A second option is to substitute each random function realization for
a discrete number, which can correspond to the scenario parameter of
Ruffo et al. \cite{rufbaz06} (where the number of geostatistical
realizations is finite and fixed, and where each different value of
the discrete parameter corresponds to a different realization).  Then,
a metamodel is fitted using this dicrete parameter as a qualitative
input variable.  However, using a metamodel is interesting when only a
few runs of the code is available, which correponds to a more limited
number of realizations of the functional input.  This restriction of
the possible realizations of the input random function to a few ones
is not appropriate in a general context.

Another strategy considers $\varepsilon(\bu)$ as an uncontrollable
parameter. A metamodel is fitted in function of the other scalar
parameters $\bX$:

\begin{equation}
  Y_m(\bX) = \ee(Y|\bX)
\end{equation}
Therefore, using the relation 

\begin{equation}\label{eqdecompvar}
  \mbox{Var}(Y) = \mbox{Var}[\ee(Y|\bX)] + \ee[\mbox{Var}(Y|\bX)] \
\end{equation}
it can be easily shown that the sensitivity indices of $Y$ given the
scalar parameters $\mathbf{X}=(X_i)_{i=1 \ldots p}$ write (Iooss et
al. \cite{ioorib07})

\begin{equation}\label{eqSiY} S_i = \frac{\mbox{Var}[\ee(Y_m|X_i)]}{\mbox{Var}(Y)},
  \quad S_{ij}=\frac{\mbox{Var}[\ee(Y_m|X_i X_j)]}{\mbox{Var}(Y)} - S_i - S_j, \quad \ldots
\end{equation} 
and can be computed by classical Monte-Carlo techniques applied on the
metamodel $Y_m$.  Therefore, using equation (\ref{eqdecompvar}), the
total sensitivity index of Y according to $\varepsilon(\bu)$
corresponds to the expectation of the unexplained part of
$\mbox{Var}(Y)$ by the metamodel $Y_m$:

\begin{equation}\label{eqsteps}
  S_{T_\varepsilon} = \frac{\ee[\mbox{Var}(Y|\bX)]}{\mbox{Var}(Y)}
\end{equation}
Using this approach, our objective is altered because it is impossible
to decompose the $\varepsilon$ effects into an elementary effect
($S_\varepsilon$) as well as the interaction effects between
$\varepsilon$ and the scalar parameters $(X_i)_{i=1 \ldots p}$.
However, we see below that our technique allows a qualitative
appraisal of the interaction indices.

The sensitivity index estimations from equations (\ref{eqSiY}) and
(\ref{eqsteps}) raise two difficulties:
\begin{enumerate}
\item It is well known that classical parametric metamodels (based on
  least squares fitting) are not adapted to estimate $\ee(Y|\bX)$
  accurately due to the presence of heteroscedasticity (induced by the
  effect of $\varepsilon$).  Such cases are analyzed by Iooss et al.
  \cite{ioorib07}. The authors show that heteroscedasticity may lead
  to sensitivity indices misspecifications.
\item Classical non parametric methods, such as Generalized Additive
  Models (Hastie and Tibshirani \cite{hastib90}) and Gaussian
  processes (Sacks et al. \cite{sacwel89}) that can provide efficient
  estimation of $\ee(Y|\bX)$ (examples are given in Iooss et al.
  \cite{ioorib07}), even in high dimensional input cases
  ($p>5$). However, these approaches are based on a homoscedasticity
  hypothesis and do not enable the estimation of $\mbox{Var}(Y|\bX)$.
\end{enumerate}

To solve the second problem, Zabalza-Mezghani et al. \cite{zabman04}
propose the use of a theory developed for experimental data (McCullagh
and Nelder \cite{macnel89}): the simultaneous fitting of the mean and
the dispersion by two interlinked Generalized Linear Models (GLM),
which is called the joint modeling (see Appendix A.1).  Besides, to
resolve the first problem, this approach has been extended by Iooss et
al.  \cite{ioorib07} to non parametric models. This generalization
allows more complexity and flexibility. The authors propose the use of
Generalized Additive Models (GAMs) based on penalized smoothing
splines (Wood \cite{woo06}).  A succint description of GAM and joint
GAM is given in Appendix A.2.  GAMs allow model and variable
selections using quasi-likelihood function, statistical tests on
coefficients and graphical display.  However, compared to other
complex metamodels, GAMs impose an additive effects hypothesis.
Therefore, two metamodels are obtained: one for the mean component
$Y_m(\bX) = \ee(Y|\bX)$; and the other one for the dispersion
component $Y_d(\bX) = \mbox{Var}(Y|\bX)$. The sensitivity indices of
$\bX$ are computed using $Y_m$ with the standard procedure
(Eq. (\ref{eqSiY})), while the total sensitivity index of
$\varepsilon(\bu)$ is computed from $\ee(Y_d)$
(Eq. (\ref{eqsteps})). Using the model for $Y_d$ as well as the
associated regression diagnostics, it is possible to deduce
qualitative sensitivity indices for the interactions between
$\varepsilon(\bu)$ and the scalar parameters of $\bX$.

One major assumption of the joint modeling approach is that the ``mean
response'' of the computer code is well handled using $Y_m$.
Consequently, all the unexplained part of the computer model by this
metamodel is due to the uncontrollable parameter. In other words, the
better the mean component metamodel is, the smaller is the influence
of the uncontrollable parameter.  This is a strong assumption which
has to be validated in order to avoid erroneous results.  In fact,
some simple statistical and graphical tools can be used while fitting
the mean component (Iooss et al. \cite{ioorib07}): the explained
deviance value, the observed responses versus predicted values plot
(and its quantile-quantile plot) and the deviance residuals plot.
This last plot allows to detect some fitting problems by revealing
possible biases or large residual values.  Some examples are given in
section \ref{secjmappl}.  These tools can also be applied for the
dispersion component fit. For a detailed overview of these diagnostic
tools, one can refer to McCullagh \& Nelder \cite{macnel89}.

\section{APPLICATION TO AN ANALYTICAL EXAMPLE}
\label{sec:analyticalExample}

The three previously proposed methods are first illustrated on a
simple analytical model with two scalar input variables and one
functional input:

\begin{equation}\label{eqWNIshi}
  Y=f(X_1,X_2,\varepsilon(t)) =
  \sin(X_1)+7\sin(X_2)^2+0.1[\max_t(\varepsilon(t))]^4\sin(X_1)
\end{equation}
where $X_i\sim \cal{U}[-\pi;\pi]$ for $i=1,2$ and $\varepsilon(t)$ is
a white noise, i.e. an i.i.d. stochastic process $\varepsilon(t)
\sim{\cal N}(0,1)$.  In our model simulations, $\varepsilon(t)$ is
discretized in one hundred values: $t=1\ldots 100$.  The function
(\ref{eqWNIshi}) is similar to the well-known Ishigami function (Homma
and Saltelli \cite{homsal96}) but substitute the third parameter for
the maximum of a stochastic process.  Consequently, we call our
function the white-noise Ishigami function (WN-Ishigami).  Although
the WN-Ishigami function is an analytical model, the introduction of
the maximum of a stochastic process inside a model is quite realistic.
For example, some computer models simulating physical phenomena can
use the maximum of time-dependent variable - river height, rainfall quantity,
temperature. Such input variable can be modeled by a temporal
stochastic process.





As for the Ishigami function, we can immediately deduce from the
formula (\ref{eqWNIshi}):

\begin{equation}\label{eqszero}
  S_\varepsilon=S_{12}=S_{2\varepsilon}=S_{12\varepsilon}=0
\end{equation}
Then, we have

\begin{equation}\label{eqst}
  S_{T_1}=S_1+S_{1\varepsilon}, \quad S_{T_2}=S_2, \quad
  S_{T_\varepsilon}=S_{1\varepsilon}
\end{equation}
In the following, we focus our attention on the estimation of $S_1$,
$S_2$ and $S_{T_\varepsilon}$.

Because of a particularly complex probability distribution of the
maximum of a white noise, there is no analytical solution for the
theoretical Sobol's indices $S_1$, $S_2$ and $S_{T_\varepsilon}$ for
the WN-Ishigami function. Even with the asymptotic hypothesis (number
of time steps tending to infinity), where the maximum of the white
noise follows Generalized Extreme Value distribution, theoretical
indices are unreachable. Therefore, our benchmark Sobol's indices
values are derived from the Monte-Carlo method.

\subsection{The macroparameter and ``trigger'' parameter methods}\label{secmacroex}

Table \ref{tab:WNIshi} contains the Sobol's index estimates using the
macroparameter and ``trigger'' parameter methods.  As explained
before, we can only use some algorithms based on independent
Monte-Carlo samples. We apply the algorithm of Sobol \cite{sob93}
that computes $S_1$, $S_2$, $S_{1\varepsilon}$ at a cost $n=4N$ and
the algorithm of Saltelli \cite{sal02} which computes the first order
indices $S_1$, $S_2$ and the total sensitivity indices $S_{T_1}$,
$S_{T_2}$, $S_{T_\varepsilon}$ at a cost $n=5N$ (where $N$ is the size
of the Monte-Carlo samples, cf. section \ref{secmacro}).  For the
estimation, the size of the Monte-Carlo samples is limited to
$N=10000$ because of memory computer limit.  Indeed, the functional
input $\varepsilon(\bu)$ contains for each simulation set $100$
values. Then, the input sample matrix has the dimension $N \times 102$
which becomes extremely large when $N$ increases.  To evaluate the
effect of this limited Monte-Carlo sample size $N$, each Sobol's index
estimate is associated to a standard-deviation estimated by bootstrap
(Saltelli et al. \cite{salrat08}) - with $100$ replicates of the
input-output sample.  The obtained standard-deviations ($sd$) are
relatively small, of the order of $0.01$, which is rather sufficient
for our exercise.

{\em Remark: We have also tried to estimate Sobol's indices with
  smaller Monte-Carlo sample sizes $N$.  The order of the obtained
  standard-deviations (estimated by bootstrap) of the Sobol's estimates
  are the following: $sd \sim 0.02$ for $N=5000$, $sd \sim 0.04$ for
  $N=1000$ and $sd \sim 0.06$ for $N=500$.  We conclude that the
  Monte-Carlo estimates are sufficiently accurate for $N > 5000$.
}

\begin{table}
  \centering
  \caption{Sobol's sensitivity indices (with standard deviations $sd$) obtained
    from two Monte-Carlo algorithms (Sobol \cite{sob93} and Saltelli \cite{sal02})
    and two integration methods of the functional input $\varepsilon$
    (macroparameter and ``trigger'' parameter)
    on the WN-Ishigami function.
    ``---'' indicates that the value is not available.} 
  \label{tab:WNIshi}
  \begin{tabular}{cccccccccccc}
    &&&&&&&&&&&\\
    \hline
    \multirow{3}*{Indices} & \multicolumn{5}{c}{Macroparameter} &&
    \multicolumn{5}{c}{``Trigger'' parameter}\\
    & \multicolumn{2}{c}{Sobol's algo} && \multicolumn{2}{c}{Saltelli' algo}
    && \multicolumn{2}{c}{Sobol's algo} && \multicolumn{2}{c}{Saltelli' algo} \\
    \cline{2-3} \cline{5-6} \cline{8-9} \cline{11-12}
    & Values & $sd$ && Values & $sd$ && Values & $sd$ && Values & $sd$\\
    \hline
    $S_1$ & 0.540 & 1.3e-2 && 0.551 & 1.6e-2 && 0.304 & 1.3e-2 && 0.330 & 1.8e-2 \\
    $S_{T_1}$ & --- & --- && 0.808 & 2.0e-2 && --- & --- && 0.656 & 1.4e-2 \\
    \\
    $S_2$ & 0.197 & 1.1e-2 && 0.207 & 0.8e-2 && 0.329 & 1.4e-2 && 0.348 & 1.5e-2 \\
    $S_{T_2}$ & --- & --- && 0.212 & 0.7e-3 && --- & --- && 0.532 & 1.3e-2 \\
    \\
    $S_{1\varepsilon}$  & 0.268 & 2.4e-2 && --- & --- && 0.177 & 2.2e-2 && --- & --- \\
    $S_{T_\varepsilon}$ & --- & --- && 0.248 & 1.3e-2 && --- & --- && 0.336 & 1.4e-2 \\
    \hline
  \end{tabular}
\end{table}

\vspace{0.5cm}
{\bf Macroparameter}

For the macroparameter method, the theoretical relations between
indices given in (\ref{eqst}) are satisfied.  We are therefore
confident with the estimates obtained with this method and we choose
the Sobol's indices obtained with Saltelli's algorithm as the reference
indices:

\begin{equation*}
  S_1=55.1\%, \quad S_2=20.7\%, \quad S_{T_\varepsilon}=24.8\%
\end{equation*}
The $S_\varepsilon$, $S_{12}$, $S_{2\varepsilon}$ and
$S_{12\varepsilon}$ indices (Eq. (\ref{eqszero})) are not reported in
table~\ref{tab:WNIshi} as estimates are negligible.

\vspace{0.5cm}
{\bf Trigger parameter}

Using the ``trigger'' parameter method, the estimates reported in
table \ref{tab:WNIshi} are quite far from the reference values.  The
inadequacies are larger than $30\%$ for all the indices, and can be
larger than $60\%$ for a few ones ($S_2$ and $S_{T_2}$).  Moreover,
the relations given in (\ref{eqst}) are not satisfied at
all. Actually, replacing the input parameter $\varepsilon(\bu)$ by
$\xi$ which governs the presence or the absence of the functional
input parameter changes the model.  When $\varepsilon$ is not
simulated, it is replaced by its mean (zero) and the WN-Ishigami
function becomes $Y=\sin(X_1)+7\sin(X_2)^2$.  Therefore, the mix of
the WN-Ishigami model and this new model perturbs the estimation of
the sensitivity indices, even those unrelated to $\varepsilon$ (like
$X_2$).  In conclusion, the obtained results are in concordance with
the expected results.

This result confirms our expectation: sensitivity indices derived from
the ``trigger'' parameter method have not the same sense that the
classical ones, i.e., the measure of the contribution of the input
parameter variability to the output variable variability.  The
sensitivity indices obtained with these two methods are unconnected
because the ``trigger'' parameter method changes the structure of the
model.

\subsection{The joint modeling approach}\label{secjmappl}

We apply now the joint modeling approach which requires an initial
input-output sample to fit the joint metamodel - the mean component
$Y_m$ and the dispersion component $Y_d$. For our application, a
learning sample size of $n=500$ was considered; i.e., $n$ independent
random samples of $(X_1,X_2,\varepsilon(\bu))$ were simulated leading
to $n$ observations for $Y$.  Joint GLM and joint GAM fitting
procedures are fully described in Iooss et al.  \cite{ioorib07}. Some
graphical residual analyses are particularly useful to check the
relevance of the mean and dispersion components of the joint models.
In the following, we give the results of the joint models fitting on a
learning sample $(X_1,X_2,\varepsilon(\bu),Y)$.  Let us recall that we
fit a model to predict $Y$ in function of $(X_1,X_2)$.

\vspace{0.5cm}
{\bf Joint GLM fitting}

For the joint GLM, a fourth order polynomial for the parametric form
of the model is considered.  Moreover, only the explanatory terms are
retained in our regression model using analysis of deviance and the
Fisher statistics.  The equation of the mean component writes:

\begin{equation}
  Y_m = 1.77 + 4.75 X_1 + 1.99 X_2^2 -0.51 X_1^3 -0.26 X_2^4 \;.
\end{equation}
The value estimates, standard-deviation estimates and Student test results
on the regression coefficients  are given in table \ref{tab:jglm}.
Residuals graphical analysis makes it also possible to appreciate the model
goodness-of-fit. 

\begin{table}
  \centering
  \caption{For the WN-Ishigami function, summary results of the joint
    GLM fitting, for the mean component $Y_m$ and the dispersion
    component $Y_d$. Estimate standard errors as well as statistics and
    p-values for the Student's test are reported.}
  \label{tab:jglm}
  \begin{tabular}{ccccc}
    &&&&\\
    \hline
    $Y_m$ &&&&\\
    \hline
    & Estimate & Std. Error & t-value & Pr(>|t|)  \\  
    (Intercept)& 1.77495  &  0.22436   & 7.911   & 1.68e-14 \\
    $X_1$         & 4.75219  &  0.16283   & 29.186  & < 2e-16 \\
    $X_2^2$    &  1.99965 &   0.14331  & 13.953  & < 2e-16 \\
    $X_1^3$    & -0.51254 &   0.02479  & -20.679 & < 2e-16 \\
    $X_2^4$    & -0.25952 &   0.01657  & -15.659 & < 2e-16 \\
    &&&&\\
    \hline
    $\log(Y_d)$ &&&&    \\
    \hline
    & Estimate & Std. Error & t-value & Pr(>|t|)  \\  
    (Intercept)   & 1.9652     & 0.1373    & 14.32   & <2e-16\\
    \hline
  \end{tabular}
\end{table}

The explained deviance of this model is $D_{expl}=73\%$.  It can be
seen that it remains $27\%$ of non explained deviance due to the model
inadequacy and/or to the functional input parameter.  The predictivity
coefficient, i.e. coefficient of determination $R^2$ computed on a
test sample, is $Q_2=70\%$. $Q_2$ is relatively coherent with the
explained deviance.

For the dispersion component, using analysis of deviance techniques,
none significant explanatory variable were found: the heteroscedastic
pattern of the data has not been retrieved.  Thus, the dispersion
component is supposed to be constant (see Table \ref{tab:jglm}); and
the joint GLM model is equivalent to a simple GLM - but with a
different fitting procedure.

\vspace{0.5cm}
{\bf Joint GAM fitting}

At present, we try to model the data using a joint GAM.  For each
component (mean and dispersion), Student test for the parametric part
and Fisher statistics for the non parametric part allow us to keep
only the explanatory terms (see Table \ref{tab:jgam}).  The resulting
model is described by the following features:

\begin{equation}
  \begin{array}{rcl}
    Y_m & = & 3.76 - 5.54 X_1 + s_1(X_1) + s_2(X_2) \;,\\
    \log(Y_d) & = & 1.05 + s_{d1}(X_1) \;,
  \end{array}
\end{equation}
where $s_1(\cdot)$, $s_2(\cdot)$ and $s_{d1}(\cdot)$ denote three
penalized spline smoothing terms.

\begin{table}
  \centering
  \caption{For the WN-Ishigami function, summary results of the joint
    GAM fitting, for the mean component $Y_m$ and the dispersion
    component $Y_d$. Estimate standard errors as well as statistics and
    p-values for the Student's test are reported. For the smoothing
    splines, the estimated degree of freedom (edf), the rank of the
    smoother and the statistics and p-values for the null hypotheses
    that each smooth term is zero are reported.} 
  \label{tab:jgam}
  \begin{tabular}{ccccc}
    &&&&\\
    \hline
    $Y_m$ &&&&\\
    \hline
    & Estimate & Std. Error & t-value & Pr(>|t|)  \\    
    (Intercept) & 3.76439 &   0.09288  & 40.53  & <2e-16 \\
    $X_1$          & -5.53920&     0.33607 & -16.48 &  <2e-16 \\
    &&&&\\
    & edf & Est.rank   &  F &  p-value    \\
    $s_1(X_1)$ & 5.656       & 8 & 151.1  & <2e-16 \\
    $s_2(X_2)$ & 8.597       & 9 & 411.4  & <2e-16 \\
    &&&&\\
    \hline
    $\log(Y_d)$  &&&&\\
    \hline
    &  Estimate  & Std. Error& t-value & Pr(>|t|)    \\
    (Intercept)   & 1.05088   & 0.07885 &  13.33  & <2e-16\\
    &&&&\\
    & edf & Est.rank   &  F &  p-value    \\
    $s_{d1}(X_1)$ & 8.781   &     9 & 36.09  & <2e-16 \\
    \hline
  \end{tabular}
\end{table}

The explained deviance of the mean component is $D_{expl}=92\%$ and
the predictivity coefficient is $Q_2=77\%$.  Therefore, the joint GAM
approach outperforms the joint GLM one. Indeed, the proportion of
explained deviance is clearly greater for the GAM model. Even if this
is obviously related to an increasing number of parameters; this is
also explained as GAMs are more flexible than GLMs.  This is
confirmed by the increase of the predictivity coefficient - from
$70\%$ to $77\%$.  Moreover, due to the GAMs flexibility, the
explanatory variable $X_1$ is identified for the dispersion component.
The interaction between $X_1$ and the functional input parameter
$\varepsilon(\bu)$ which governs the heteroscedasticity of this model
is therefore retrieved.

\begin{figure}[ht]
  \centering
  \includegraphics[angle=-90,width=11cm]{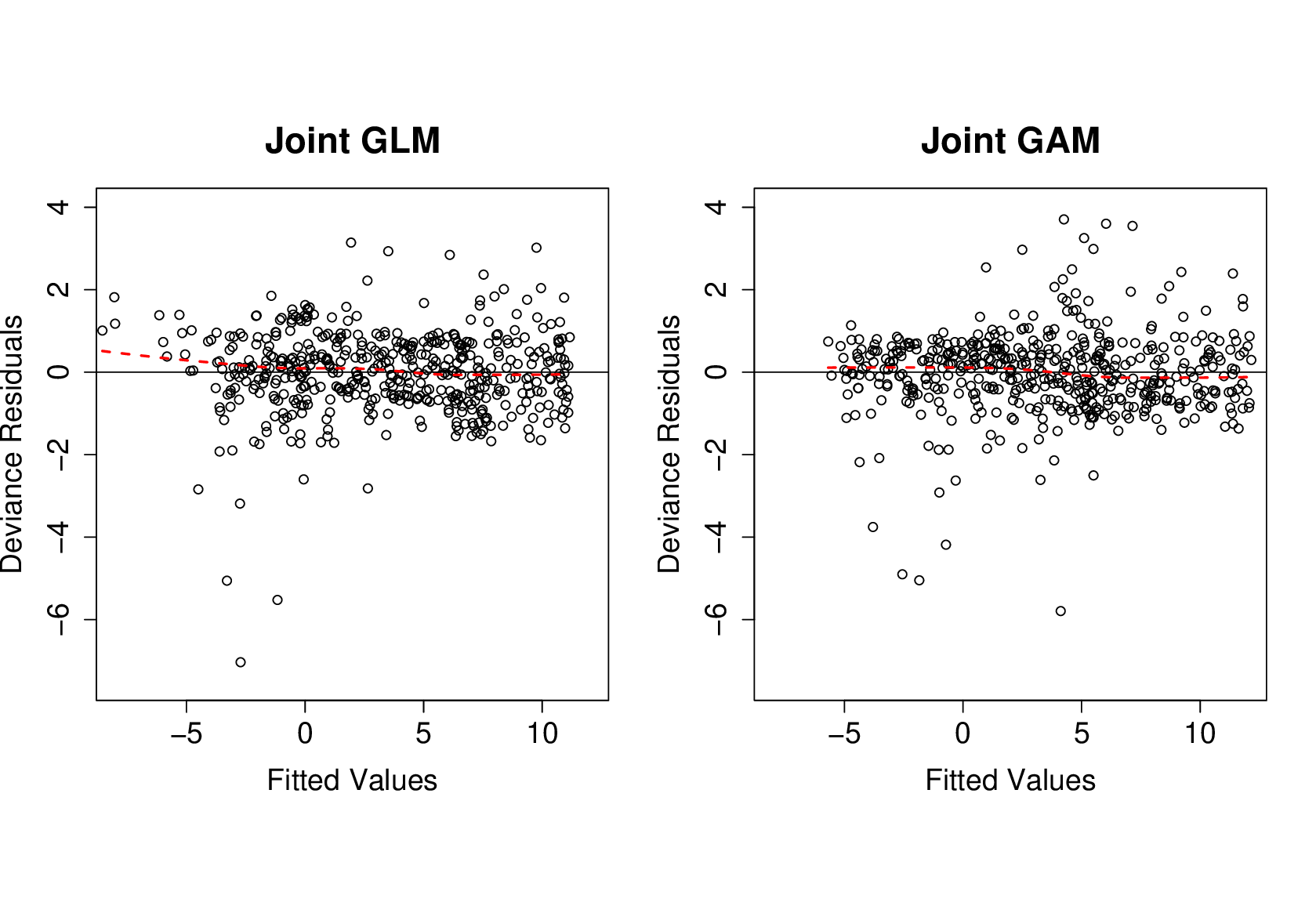}
  \caption{Deviance residuals for the joint GLM and the
    Joint GAM versus the fitted values (WN-Ishigami application).
    Dashed lines correspond to local polynomial smoothers.}
  \label{fig:devResIshigWN}
\end{figure}

Figure~\ref{fig:devResIshigWN} shows that the deviance residuals for
the mean component of the joint GAM seem to be more homogeneously
dispersed around the $x$-axis than the deviance residuals of the joint
GLM.  This leads to a better prediction from the joint GAM on the
whole range of the observations.

\vspace{0.5cm}
{\bf Sobol's indices}

From the joint GLM and the joint GAM, Sobol's sensitivity indices can
be computed using equations (\ref{eqSiY}) and (\ref{eqsteps}) - see
Table \ref{tab:WNIshiSobol}. The reference values are extracted from
the results of the macroparameter method via Saltelli's algorithm (see
Table~\ref{tab:WNIshi}) and from the WN-Ishigami analytical form
(\ref{eqWNIshi}) (for example we know that $S_{12}=0$ because there is
no interaction between $X_1$ and $X_2$).  The standard deviation
estimates ($sd$) are obtained from $100$ replicates of the Monte-Carlo
estimation procedure - which uses $N=10000$ for the size of the
Monte-Carlo samples (see section \ref{secmacro}).  The joint GLM and
joint GAM give approximately good estimates of $S_1$ and
$S_2$. Despite the joint GLM leads to an acceptable estimation for
$S_{T_\varepsilon}$, we will see later that it is fortuitous.  The
estimation of $S_{T_\varepsilon}$ with the joint GAM seems also
satisfactory but not accurate.  In fact, an efficient modeling of
$\mbox{Var}(Y|\bX)$ is difficult, which is a common statistical
difficulty in heteroscedastic regression problems (Antoniadis \&
Lavergne \cite{antlav95}).

Another way to estimate the total sensitivity index
$S_{T_\varepsilon}$ is to compute the unexplained variance of the mean
component model given directly by $1-Q_2$, where $Q_2$ is the
predictivity coefficient of the mean component model. In practical
applications, $Q_2$ can be estimated using leave-one-out or cross
validation procedures. In our analytical case, the index is estimated
with the former method and leads to a correct estimation - $0.23$
instead of $0.25$.

\begin{table}[ht]
  \centering
  \caption{Sobol's sensitivity indices (with standard deviations) for
    the WN-Ishigami function: exact and estimated values from joint
    GLM and joint GAM (fitted with a $500$-size sample). ``Method''
    indicates the estimation method:   MC for the Monte-Carlo
    procedure, Eq for a deduction from the model equations and $Q_2$
    for the deduction of the predictivity coefficient $Q_2$. ``---''
    indicates that the value is not available.} 
  \label{tab:WNIshiSobol}
  \begin{tabular}{clcclccclccc}
    &&&&&&&&&&\\
    \hline
    \multirow{2}*{Indices} && \multicolumn{2}{c}{Reference} &&
    \multicolumn{3}{c}{Joint GLM} && \multicolumn{3}{c}{Joint GAM} \\
    \cline{3-4} \cline{6-8} \cline{10-12}
    && Values & $sd$ && Values & $sd$ & Method && Values & $sd$ & Method \\
    \hline
    $S_1$ && 0.551 & 1.6e-2                && 0.580 & 3e-3 & MC && 0.554 & 4e-3 & MC \\
    $S_2$ && 0.207 & 0.8e-2                && 0.181 & 7e-3 & MC && 0.228 & 6e-3 & MC \\
    $S_{T_\varepsilon}$ && 0.248 & 1.3e-2  && 0.268 & 1e-3 & MC && 0.218 & 1e-3 & MC \\
    && &               && 0.30 & --- & $Q_2$ && 0.23 & --- & $Q_2$ \\
    $S_{12}$ && 0 &                        && 0 & --- & Eq && 0 & --- & Eq  \\
    $S_{1\varepsilon}$ && 0.248 & 1.3e-2   && 0 & --- & Eq && $]0,0.23]$ & --- & Eq \\
    $S_{2\varepsilon}$ && 0 &              && 0 & --- & Eq && 0 & --- & Eq \\
    $S_{12\varepsilon}$ && 0 &             && 0 & --- & Eq && 0 & --- & Eq  \\
    $S_{T_1}$ && 0.808 & 2.0e-2            && 0.580 & 3e-3 & Eq && $]0.554,0.784]$ & --- & Eq \\
    $S_{T_2}$ && 0.212 & 0.7e-3            && 0.181 & 7e-3 & Eq && 0.228 & 6e-3 & Eq \\ 
    $S_\varepsilon$ && 0        &          && 0.268 & 1e-3 & Eq && $[0,0.23]$ & --- & Eq \\
    \hline
  \end{tabular}
\end{table}

For the other sensitivity indices, the conclusions draw from the GLM
formula are completely erroneous. As the dispersion component is
constant, $S_\varepsilon=S_{T_\varepsilon}=0.268$ while
$S_\varepsilon=0$ in reality. In contrary, the deductions draw from
GAM formulas are exact: $(X_1,\varepsilon)$ interaction sensitivity is
strictly positive ($S_{1\varepsilon}>0$) because $X_1$ is active in
the dispersion component $Y_d$,
$S_{2\varepsilon}=S_{12\varepsilon}=0$, $S_{T_2}=S_2$ and
$S_{12}=S_{23}=S_{123}=0$.  The drawback of this method is that some
indices ($S_{1\varepsilon}$, $S_{\varepsilon}$ and $S_{T_1}$) remain
unknown due to the non separability of the dispersion component
effects.  However, we can easily deduce some variation intervals which
contain these indices: $S_\varepsilon$ and $S_{1\varepsilon}$ are
smaller than $S_{T_\varepsilon}$ while $S_1+\min(S_{1\varepsilon})
\leq S_{T_1} \leq S_1+\max(S_{1\varepsilon})$.  All these additional
information provide qualitative importance measures for the unknown
indices.

\begin{figure}[ht]
  \centering
  \includegraphics[width=\textwidth]{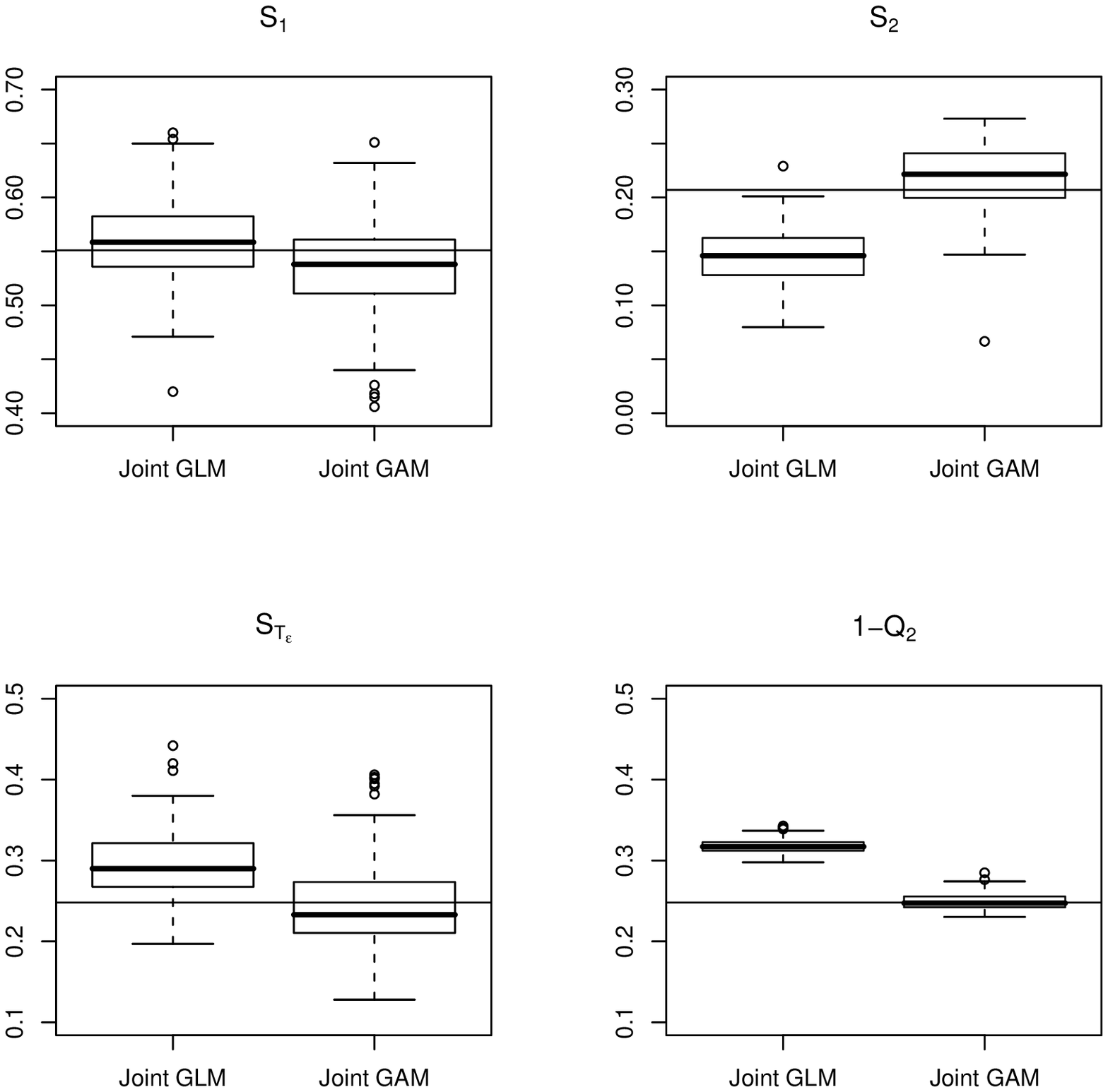}
  \caption{WN-Ishigami application. Comparison of Sobol's indices
    estimates for the learning sample size: $n=200$.
    For each index, the horizontal line is the reference value.}
  \label{fig:IshigamiN200}
\end{figure}

\begin{figure}[ht]
  \centering
  \includegraphics[width=\textwidth]{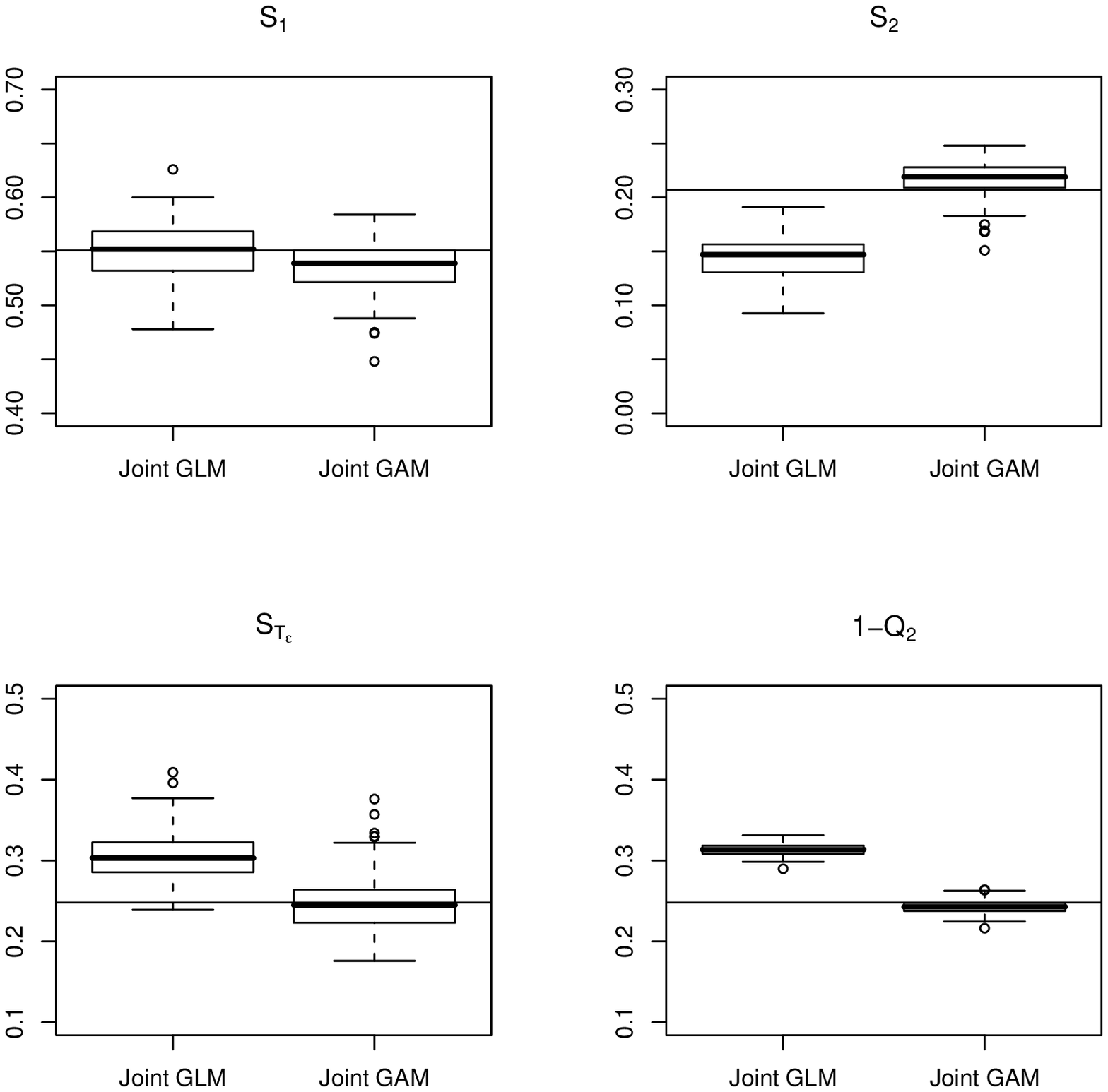}
  \caption{WN-Ishigami application. Comparison of Sobol's indices
    estimates for the learning sample size: $n=500$.
    For each index, the horizontal line is the reference value.}
  \label{fig:IshigamiN500}
\end{figure}

By estimating Sobol's indices with those obtained from other learning
samples, we observe that the estimates are rather dispersed: it seems
that the estimates are not robust according to different learning
samples for the joint models.  To examine this effect, we propose to
study two different sample sizes ($n=200$ and $n=500$).  For each
sample size, the distribution of the Sobol's indices estimates is
assessed using a bootstrap procedure.  Figures \ref{fig:IshigamiN200}
and \ref{fig:IshigamiN500} show the results of this investigation,
which are particularly convincing.  Several conclusion can be drawn:
\begin{itemize}
\item For the joint GAM, the boxplot interquartile interval of each
  index contains its reference value.  In contrary, the joint GLM
  fails to obtain correct estimates: except for $S_1$, the sensitivity
  reference values are outside the interquartile intervals of the
  obtained boxplots.
\item The superiority of the joint GAM with respect to the joint GLM
  is corroborated, especially for $S_2$ and $S_{T_\varepsilon}$.
\item The increase of the learning sample size has no effect on the
  joint GLM results (due to the parametric form of this model).
  However, for the joint GAM, boxplots widths are strongly reduced
  from $n=200$ to $n=500$.  In addition, the mean estimates seem to
  converge to the reference values.
\item As explained before, the estimation of $S_{T_\varepsilon}$ using
  the predictivity coefficient $Q_2$ is markedly better than through
  the dispersion component model. This is not the case for the joint
  GLM. Moreover, we confirm that the previous result of table
  \ref{tab:WNIshiSobol}, $S_{T_\varepsilon}=0.268$, was a good case:
  with $100$ replicates, $S_{T_\varepsilon}$ ranges from $0.24$ to
  $0.35$ (Figure \ref{fig:IshigamiN500}).
\end{itemize}


In conclusion, this example shows that the joint models, and specially
the joint GAM, can adjust rather complex heteroscedastic situations.
Of course, additional tests are needed to confirm this result.
Moreover, the joint models offer a theoretical basis to compute
efficiently global sensitivity indices for models with functional
input parameter.  Finally, the required number of computer model
evaluations is much smaller with the joint modeling method (here
$n=200$ or $n=500$ gives good results with the joint GAM) compared to
the one of Monte-Carlo based techniques.  For exemple, using the
macroparameter method (cf. section \ref{secmacroex}) and taking
$N=5000$, we need to compute $n=25000$ model evaluations to estimate
first order and total sensitivity indices (via Saltelli's algorithm).

\section{APPLICATION TO A NUCLEAR FUEL IRRADIATION SIMULATION}

The METEOR computer code, developed within the Fuel Studies Department
in CEA Cadarache, studies the thermo-mechanical behavior of the fuel
rods under irradiation in a nuclear reactor core.  In particular, it
computes the fission gas swelling and the cladding creep (Garcia et
al. \cite{garstr02}). These two output variables are considered in our
analysis.  These variables are of fundamental importance for the
physical understanding of the fuel behavior and for the monitoring of
the nuclear reactor core.

Input parameters of such mechanical models can be evaluated either by
database analyses, arguments invoking simplifying hypotheses, expert
judgment. All these considerations lead to assign to each input
parameter a nominal value associated with an uncertainty.  In this
study, six uncertain input parameters are considered: the initial
internal pressure $X_1$, the pellet and cladding radius $X_2$, $X_3$,
the microstructural fuel grain diameter $X_4$, the fuel porosity $X_5$
and the time-dependent irradiation power $P(t)$.  $X_1,\; \ldots, \;
X_5$ are all modeled by Gaussian independent random variables with the
following coefficient of variations: $\mbox{cv}(X_1)=0.019$,
$\mbox{cv}(X_2)=1.22\times10^{-3}$,
$\mbox{cv}(X_3)=1.05\times10^{-3}$, $\mbox{cv}(X_4)=0.044$,
$\mbox{cv}(X_5)=0.25$.  The last variable $P(t)$ is a temporal
function (discretized in $3558$ values) and its uncertainty
$\varepsilon(t)$ is modelled like a stochastic process.  For
simplicity, an Additive White Noise (AWN), of uniform law ranging
between $-5\%$ and $+5\%$, was introduced.

As in the previous application, additionally to its scalar random
variables, the model includes an input functional variable $P(t)$.  To
compute Sobol's indices of this model, we have first tried to use the
macroparameter method.  We have succedeed to perform the calculations
with $N=1000$ (for the Monte-Carlo sample sizes of
Eqs. (\ref{eqsobolalgo1}), (\ref{eqsobolalgo2}) and
(\ref{eqsobolalgo3})).  The sensitivity indices estimates have been
obtained after $10$ computation days and were extremely imprecise,
with strong variations between $0$ and $1$.  Because of the required
cpu time, an increase of the sample size $N$ to obtain acceptable
sensitivity estimates was unconceivable.
Therefore, the goal of this section is to show how the use of the
joint modeling approach allows to estimate the sensitivity
indices of the METEOR model and, in particular, to quantify the 
functional input variable influence.

$500$ METEOR calculations were carried out using Monte-Carlo sampling
of the input parameters (using Latin Hypercube Sampling).  As
expected, the AWN on $P(t)$ generates an increase in the standard
deviation of the output variables (compared to simulations without a
white noise): $6\%$ increase for the variable \emph{fission gas
  swelling} and $60\%$ for the variable \emph{cladding creep}.

\subsection{Gas swelling}

We start by studying the gas swelling model output.  With a joint GLM,
the following result for $Y_m$ and $Y_d$ were obtained:

\begin{equation}\label{eqgas}
  \begin{cases}
    Y_m & = -76 -0.4X_1 +20X_2 + 8X_4 + 134X_5 + 0.02X_4^2 - 2X_2 X_4
    - 6 X_4 X_5\\ 
    \log(Y_d) & = -2.4 X_1
  \end{cases}
\end{equation}
The explained deviance of the mean component is $D_{expl}=86\%$.  As
the residual analyses of mean and dispersion components do not show
any biases, the resulting model seems satisfactory.
The joint GAM was also fitted on these data and led to similar
results. Thus, it seems that spline terms are useless and that a joint
GLM model is appropriate.

Table \ref{tab:meteor} shows the results for the Sobol's indices
estimation using Monte-Carlo methods applied on the metamodel (\ref{eqgas}).  
The standard deviation ($sd$)
estimates are obtained from $100$ replicates of the Monte-Carlo
estimation procedure -which uses $10^5$ model computations for one
index estimation.  It is useless to perform the Monte-Carlo
estimation for some indices because they can be deduced from the
joint model equations.  For example, $S_3=0$ (resp. $S_{\varepsilon
  2}=0$) because $X_3$ (res. $X_2$) is not involved in the mean
(resp. dispersion) component in equation (\ref{eqgas}).  Moreover,
we know that $S_{1 \varepsilon} >0$ because $X_1$ is an explanatory
variable inside the dispersion component $Y_d$.
However, this formulation does not allow to have any idea about $S_\epsilon$
which reflects the first order effect of $\varepsilon$.
Therefore, some indices are not accessible, such as
$S_\varepsilon$ and $S_{1 \varepsilon}$ non distinguishable inside
the total sensitivity index $S_{T_\varepsilon}$.
Finally, we can check that $\displaystyle \sum_{i=1}^5 S_i +
\sum_{i,j=1,i<j}^5 S_{ij} + S_{T_\varepsilon} = 1$ holds - up to
numerical approximations.

It can be seen that $X_4$ (grain diameter) and $X_5$ (fuel porosity)
are the most influent factors (each one having $40\%$ of influence),
and do not interact with the irradiation power $P(t)$ (represented
by its uncertainty $\varepsilon$).  In addition, the effect of $P(t)$
is not negligible ($S_{T_\varepsilon}=14\%$) and parameter $X_1$ (internal pressure)
acts mainly with its interaction with $P(t)$.
A sensitivity analysis by fixing $X_1$ could allow us to obtain
some information about the first order effect of $\varepsilon$
in the model.

\subsection{Cladding creep}

We study now the cladding creep model output.
With a joint GLM, the model for $Y_m$ and $Y_d$ is:

\begin{equation}
  \begin{cases}
    Y_m & = -2.75 + 1.05 X_2 - 0.15 X_3 - 0.58 X_5\\
    \log(Y_d) & = 156052 - 76184 X_2 + 9298 X_2^2
  \end{cases}
\end{equation}
The explained deviance of the mean component is $D_{expl}=26\%$.  As
the residual analyses of mean and dispersion components show some
biases, the resulting model is not satisfactory.

For the joint GAM, the spline terms $\lbrace s(X_2), s(X_3), s(X_5)
\rbrace$ and $s(X_2)$ are added within the mean component and the
dispersion component respectively.  The explained deviance of the mean
component is $D_{expl}=29\%$ which is not significantly greater than
$26\%$. However, as the mean component residual biases of the joint
GAM are smaller than those observed for the joint GLM, the joint GAM
seems to be more relevant than the joint GLM.

Table \ref{tab:meteor} shows the Sobol's index estimates using
Monte-Carlo methods and deductions from the joint model equations.
For the joint GLM and joint GAM of the cladding creep, $\displaystyle
\sum_{i=1}^5 S_i + \sum_{i,j=1,i<j}^5 S_{ij} + S_{T_\varepsilon}=1$
holds -- up to numerical imprecisions.  Due to the proximity of the
two joint models, results are similar.
This analysis shows that the parameter $X_2$ (pellet radius) explains
$28\%$ of the uncertainty of the cladding creep phenomenon, while the
other scalar parameters have negligible influence. The greater part of
the cladding creep variance ($S_{T_\varepsilon}=70\%$) is explained by the irradiation
power uncertainty (the AWN). Physicists may be interested in quantifying the
interaction influence between the pellet radius and the irradiation
power. Unfortunately, this interaction is not available for the moment
in our analysis.


\begin{table}
  \caption{Sobol's sensitivity indices (with standard deviations $sd$) from
    joint models fitted on the outputs of the METEOR code. ``Method''
    indicates the estimation method: MC for the Monte-Carlo procedure
    and Eq for a deduction from the joint model equation. 
    ``---'' indicates that the value is not available.} 
  \label{tab:meteor}
  \hspace{-2cm}
  \begin{tabular}{cccccccccccc}
    &&&&&&&&&&&\\
    \hline
    \multirow{3}*{Indices} & \multicolumn{3}{c}{Gas swelling} &&
    \multicolumn{7}{c}{Cladding creep}\\
    & \multicolumn{3}{c}{Joint GLM} && \multicolumn{3}{c}{Joint GLM}
    && \multicolumn{3}{c}{Joint GAM} \\
    \cline{2-4} \cline{6-8} \cline{10-12}
    & Values & $sd$ & Method && Values & $sd$ & Method && Values & $sd$
    & Method\\
    \hline
    $S_1$ & 0.029 & 6e-3 & MC && 0.000 & 1e-3 & MC && 0.000 & 1e-3 &
    MC \\
    $S_2$ & 0.024 & 5e-3 & MC && 0.294 & 1e-4 & MC && 0.282 & 2e-4 &
    MC \\
    $S_3$ & 0 & --- & Eq && 0.006 & 1e-3 & MC && 0.007 & 1e-3 & MC \\
    $S_4$ & 0.394 & 5e-3 & MC && 0.000 & 1e-3 & MC && 0.000 & 1e-3 & MC
    \\ 
    $S_5$ & 0.409 & 6e-3 & MC && 0.006 & 1e-3 & MC && 0.006 & 1e-3 &
    MC \\ 
    $S_{24}$ & 0.002 & 5e-3 & MC &&  0 & --- & Eq &&  0 & --- & Eq
    \\
    $S_{45}$ & 0.000 & 9e-3 & MC &&  0 & --- & Eq &&  0 & --- & Eq
    \\ 
    other $S_{ij}$ & 0 & --- & Eq &&  0 & --- & Eq &&  0 & --- & Eq
    \\  
    $S_{T_\varepsilon}$ & 0.143 & 1e-4 & MC && 0.694 & 1e-4 & MC &&
    0.704 & 3e-4 & MC \\ 
    $S_{\varepsilon}$ & $[0,0.143]$ & --- & --- && $[0,0.694]$ & --- & --- && $[0,0.704]$ &
    --- & ---\\
    $S_{1\varepsilon}$ & $]0,0.143]$ & --- & --- && 0 & --- & Eq && 0 & --- &
    Eq \\
    $S_{2\varepsilon}$ & 0 & --- & Eq && $]0,0.694]$ & --- & --- && $]0,0.704]$ & ---
    & ---\\
    other $S_{i \varepsilon}$ & 0 & --- & Eq && 0 & --- & Eq &&  0 &
    --- & Eq \\
    $S_{T_1}$ & $]0.029,0.172]$ & --- & --- && 0.000  & 1e-3 & Eq &&  0.000 &
    4e-3 & Eq\\ 
    $S_{T_2}$ & 0.026 & 7e-3 & Eq && $]0.294,0.988]$ & --- & --- && $]0.282,0.986]$ & --- &
    --- \\
    $S_{T_3}$ & 0 & --- & Eq &&  0.006 & 1e-3 & Eq && 0.007 & 4e-3 &
    Eq \\  
    $S_{T_4}$ & 0.396 & 7e-3 & Eq && 0.000 & 1e-3 & Eq && 0.000 &
    4e-3 & Eq \\  
    $S_{T_5}$ & 0.409 & 0.011 & Eq && 0.006 & 1e-3 & Eq && 0.006 &
    4e-3 & Eq \\ 
    \hline
  \end{tabular}
\end{table}

\section{CONCLUSION}

This paper has proposed a solution to perform global sensitivity
analysis for time consuming computer models which depend on functional
input parameters, such as a stochastic process or a random field.  Our
purpose concerned the computation of variance-based importance
measures of the model output according to the uncertain input
parameters.  We have discussed a first natural solution which consists
in integrating the functional input parameter inside a macroparameter,
and using standard Monte-Carlo algorithms to compute sensitivity
indices. This solution is not applicable for time consuming computer
code.  We have discussed another solution, used in previous studies,
based on the replacement of the functional input parameter by a
``trigger'' parameter that governs the integration or not of the
functional input uncertainties.
However, the estimated sensitivity indices are not the expected ones
due to changes in the model structure carrying out by the method
itself.  Finally, we have proposed an innovative strategy, the joint
modeling method, based on a preliminary step of double (and joint)
metamodel fitting, which resolves the large cpu time problem of
Monte-Carlo methods.  It consists in rejecting the functional input
parameters in noisy input variables.  Then, two metamodels depending
only on the scalar random input variables are simultaneously fitted:
one for the mean function and one for the dispersion (variance)
function.

Tests on an analytical function have shown the relevance of the joint
modeling method, which provides all the sensitivity indices of the
scalar input parameters and the total sensitivity index of the
functional input parameter.  In addition, it reveals in a qualitative
way the influential interactions between the functional parameter and
the scalar input parameters.  It would be interesting in the future to
be able to distinguish the contributions of several functional input
parameters that are currently totally mixed in one sensitivity index.
This is the main drawback of the proposed method in its present form.

In an industrial application, the usefulness and feasibility of our
methodology has been established.  Indeed, other methods are not
applicable in this application because of large cpu time of the
computer code.  To a better understanding of the model behavior, the
information brought by the global sensitivity analysis can be very
useful to the physicist or the modeling engineer.  The joint model can
also be useful to propagate uncertainties in complex models,
containing input random functions, to obtain some mean predictions
with their confidence intervals.

\section{ACKNOWLEDGMENTS}

This work was supported by the ``Simulation'' program managed by the
CEA/Nuclear Energy Division. We are grateful to P. Obry (CEA
Cadarache/Département d'Etude des Combustibles) for the authorization
to use the METEOR application.  All the statistical parts of this work
have been performed within the R environment and the ``sensitivity''
and ``JointModeling'' packages.  We are grateful to the referees whose
comments significantly helped to improve the paper.

\section*{APPENDIX A: JOINT MODELING OF MEAN AND DISPERSION}

\subsection*{A.1 Joint Generalized Linear Models}

GLMs allow to extend traditional linear models by the use of a
distribution which belongs to the exponential family and a link
function that connects the explanatory variables to the explained
variable (Nelder \& Wedderburn \cite{nelwed72}).  The joint GLM
consists in putting a GLM on the mean component of the model and a GLM
on the dispersion component of the model.

The mean component is therefore described by:

\begin{equation}\label{eqGlmM}
  \begin{cases}
    \mbox{E}(Y_i) & = \mu_i,\qquad \eta_i = g(\mu_i) = \sum_j
    x_{ij}\beta_j \;,\\ 
    \mbox{Var}(Y_i) &= \phi_i v(\mu_i) \;,
  \end{cases}
\end{equation}
where $(Y_i)_{i=1\ldots n}$ are independent random variables with mean
$\mu_i$; $x_{ij}$ are the observations of the parameter $X_j$;
$\beta_j$ are the regression parameters that have to be estimated;
$\eta_i$ is the mean linear predictor; $g(\cdot)$ is a differentiable
monotonous function (called the link function); $\phi_i$ is the
dispersion parameter and $v(\cdot)$ is the variance function.  To
estimate the mean component, the functions $g(\cdot)$ and $v(\cdot)$
have to be specified.  Some examples of link functions are the
identity (traditional linear model), root square, logarithm, and
inverse functions. Some examples of variance functions are the
constant (traditional linear model), identity and square functions.

Within the joint modeling framework, the dispersion parameter $\phi_i$
is not supposed to be constant as in a traditional GLM, but is
supposed to vary according to the model:

\begin{equation}\label{eqGlmD}
  \begin{cases}
    \mathbb{E}(d_i) &= \phi_i,\qquad \zeta_i = h(\phi_i) = \sum_j
    u_{ij} \gamma_j \;,\\ 
    \mbox{Var}(d_i) &= \tau v_d(\phi_i) \;,
  \end{cases}
\end{equation}
where $d_i$ is a statistic representative of the dispersion,
$\gamma_j$ are the regression parameters that have to be estimated,
$h(\cdot)$ is the dispersion link function, $\zeta_i$ is the
dispersion linear predictor, $\tau$ is a constant and $v_d(\cdot)$ is
the dispersion variance function.  $u_{ij}$ are the observations of
the explanatory variable $U_j$.  The variables $(U_j)$ are generally
taken among the explanatory variables of the mean $(X_j)$, but can
also be different.  To ensure positivity, a log link function is often
chosen for the dispersion component.  For the statistic representing the
dispersion $d$, the deviance contribution (which is close to the
distribution of a $\chi^2$) is considered.  Therefore, as the $\chi^2$
is a particular case of the Gamma distribution, $v_d(\phi)=\phi^2$ and
$\tau \sim 2$.  In particular, for the Gaussian case, these relations
are exact: $d$ is $\chi^2$ distributed and $\tau=2$.

The joint model is fitted using Extended Quasi-Loglikelihood (EQL)
(Nelder \& Pregibon \cite{nelpre87}) maximization. The EQL behaves as
a log-likelihood for both mean and dispersion parameters.
This justifies an iterative procedure to fit the joint model. 
Statistical tools available in the GLM fitting are also available for
each component of the joint model: deviance analysis, Student and
Fisher tests, residuals graphical analysis. It allows to make some
variable selection in order to simplify model expressions.

\subsection*{A.2 Joint Generalized Additive Models}

Generalized Additive models (GAM) allow the linear term in the
linear predictor $\eta=\sum_j \beta_j X_j$ of equation (\ref{eqGlmM})
to be replaced by a sum of smooth functions $\eta=\sum_j s_j(X_j)$ (Hastie \&
Tibshirani \cite{hastib90}).
The $s_j(.)$'s are unspecified functions that are obtained by fitting
a smoother to the data, in an iterative procedure.  GAMs provide a
flexible method for identifying nonlinear covariate effects in
exponential family models and other likelihood-based regression
models.  The fitting of GAM introduces an extra level of iteration in
which each spline is fitted in turn assuming the others known.  GAM
terms can be mixed quite generally with GLM terms in deriving a model.

One common choice for $s_j$ are the smoothing splines, i.e. splines
with knots at each distinct value of the variables.  In regression
problems, smoothing splines have to be penalized in order to avoid
data overfitting.  Wood \cite{woo06} has described in details how GAMs
can be constructed using penalized regression splines.  This approach
is particularly appropriate as it allows the integrated model
selection using Generalized Cross Validation (GCV) and related
criteria, the incorporation of multi-dimensional smooths and
relatively well founded inference using the resulting models.  Because
numerical models often exhibit strong interactions between input
parameters, the incorporation of multi-dimensional smooth (for example
the bi-dimensional spline term $s_{ij}(X_i,X_j)$) is particularly
important in our context.

GAMs are generally fitted using penalized likelihood maximization. For
this purpose, the likelihood is modified by the addition of a penalty
for each smooth function, penalizing its ``wiggliness''. Namely, the
penalized loglikelihood (PL) is defined as:

\begin{equation}
  \label{eq:pllik}
  PL = L + \sum_{j=1}^p \lambda_j \int \left(\frac{\partial^2
      s_j}{\partial x_j^2} \right)^2 dx_j
\end{equation}
where $L$ is the loglikelihood function, $p$ is the total number of
smooth terms and $\lambda_j$ are ``tuning'' constants that compromise
between goodness of fit and smoothness.  Estimation of these
``tuning'' constants is generally achieved using the GCV score
minimization (Wood \cite{woo06}).

We have seen that GAMs extend in a natural way GLMs.  Iooss et
al. \cite{ioorib07} have shown how to extend joint GLM to joint GAM.
Extension of PL to penalized extended quasi-likelihood (PEQL) is
straightforward by substituting the likelihood function $PL$ and the
deviance $d$ for their extended quasi counterparts.  The fitting
procedure of the joint GAM is similar to the one of joint GLM.


\singlespacing
\bibliographystyle{plain}
\bibliography{bibl_hal}

\end{document}